\documentstyle[12pt,graphicx]{article}
\begin{document}
\title{Instantons in 4-dimensional gauged O(5) Skyrme models}
\author{{\large Y. Brihaye}$^{\ddagger}$,
{\large V. Paturyan}$^{\dagger}$,
{\large B.M.A.G Piette}$^{\diamond}$
and {\large D. H. Tchrakian}$^{\dagger \star}$ \\ \\
$^{\ddagger}${\small Physique-Math\'ematique, Universite de 
Mons-Hainaut, Mons, Belgium}\\ \\
$^{\diamond}${\small  Department of Mathematical Sciences, University of 
Durham,}\\
{\small Durham DH1 3LE United-Kingdom}\\ \\
$^{\dagger}${\small Department of
Mathematical Physics, National University of Ireland Maynooth,} \\
{\small Maynooth, Ireland} \\ \\
$^{\star}${\small School of Theoretical Physics -- DIAS, 10 Burlington Road,
Dublin 4, Ireland }}

\date{}
\newcommand{\dd}{\mbox{d}}\newcommand{\tr}{\mbox{tr}}
\newcommand{\la}{\lambda}
\newcommand{\ka}{\kappa}
\newcommand{\al}{\alpha}
\newcommand{\ga}{\gamma}
\newcommand{\de}{\delta}
\newcommand{\si}{\sigma}
\newcommand{\ee}{\end{equation}}
\newcommand{\eea}{\end{eqnarray}}
\newcommand{\be}{\begin{equation}}
\newcommand{\bea}{\begin{eqnarray}}
\newcommand{\ii}{\mbox{i}}\newcommand{\e}{\mbox{e}}
\newcommand{\pa}{\partial}\newcommand{\Om}{\Omega}
\newcommand{\vep}{\varepsilon}
\newcommand{\bfph}{{\bf \phi}}
\newcommand{\lm}{\lambda}
\def\theequation{\arabic{equation}}
\renewcommand{\thefootnote}{\fnsymbol{footnote}}
\newcommand{\re}[1]{(\ref{#1})}
\newcommand{\R}{{\rm I \hspace{-0.52ex} R}}
\newcommand{\N}{{\sf N\hspace*{-1.0ex}\rule{0.15ex}%
{1.3ex}\hspace*{1.0ex}}}
\newcommand{\Q}{{\sf Q\hspace*{-1.1ex}\rule{0.15ex}%
{1.5ex}\hspace*{1.1ex}}}
\newcommand{\C}{{\sf C\hspace*{-0.9ex}\rule{0.15ex}%
{1.3ex}\hspace*{0.9ex}}}
\newcommand{\eins}{1\hspace{-0.56ex}{\rm I}}
\renewcommand{\thefootnote}{\arabic{footnote}}

\maketitle
\begin{abstract}
We consider a family of four-dimensional non-linear sigma models
based on an O(5) symmetric group, whose fields take their values on
the 4-sphere $S^4$. An SO(4)-subgroup of the model is gauged.
The solutions of the model are characterised by two distinct topological
charges, the Chern-Pontryagin charge of the gauge field and the degree
of the map, i.e. the winding number, of the $S^4$ field. The one
dimensional equations arising from the variation of the action density
subjected to spherical symmetry are integrated numerically. Several
properties of the solutions thus constructed are pointed out. The only
solution with {\it unit} Chern-Pontryagin charge are the usual  BPST
instantons with zero $S^4$ winding number, while solutions with
{\it unit} $S^4$ winding number have zero Chern-Pontryagin charge.
\end{abstract}
\medskip
\medskip
\newpage

\section{Introduction}
\label{introduction}

The model considered in this work is described by the Lagrangian
on $\R^4$
\bea
{\cal L}=\frac{\lambda_0}{24} |F_{\mu \nu}^{[\alpha \beta]}|^2
&+&\frac{\lambda_1}{2} |D_{\mu}\phi^a|^2 
+\frac{\lambda_2}{24} |D_{\mu}\phi^a \times D_{\nu}\phi^b|^2
\nonumber \\ 
&+&\frac{\lambda_3}{72}
|D_{\mu}\phi^a \times D_{\nu}\phi^b \times  D_{\rho}\phi^c |^2
+V(\phi^5,\cos\omega)\ ,
\label{L}  
\eea
in terms of the $S^4$ valued fields $\phi^a=(\phi^{\alpha},\phi^5)$,
($\alpha$=1,2,3,4) satisfying the constraint
\[
\phi^a\phi^a\ =\ 1\ ,
\]
and the $SO(4)$ gauge connection
$A_{\mu}^{[\alpha \beta]}$ with curvature $F_{\mu \nu}^{[\alpha \beta]}$.
The covariant derivatives in \re{L} are defined by
\be
\label{cov}
D_{\mu}\phi^{\alpha}=\pa_{\mu}\phi^{\alpha}+
A_{\mu}^{[\alpha \beta]}\phi^{\beta}\ , \quad D_{\mu}\phi^5=
\pa_{\mu}\phi^5\ .
\ee
(The brackets $[..]$ imply antisymmetrisation of indices throughout.)

The Lagrangian \re{L} differs from that of the various models considered
in \cite{nl}, in the physically important\footnote{Without a {\it
quadratic} kinetic term, it is not possible to infer from {\it finite
action conditions}, that that the matter field (in this case the $S^4$
valued field) becomes asymptotically a constant and hence consistent
with {\it vacuum} field. It would then be impossible to interpret the
resulting topologically stable finite action solution as an instanton.}
respect that the kinetic term
{\em quadratic} in the $S^4$ field $\phi^a$, which was absent
there~\cite{nl}, is present here. As a result of the Derrick scaling
requirement, the system \re{L} features also a kinetic term {\em sextic}
in the $S^4$ valued field. \re{L} differs from that in \cite{nl} also in
the presence of the generic potential $V(\phi^5,\cos\omega)$
\be
\label{pott}
V(\phi^5,\omega)=\lambda(\cos \omega-\phi^5)^n\ ,\quad
0\le\omega\le\pi\ , \quad n={\rm integer}\ ,
\ee
whose role it is
to fix the asymptotic value of the field $\phi^a$, rather like the
pion--mass potential in the usual 3 dimensional $O(4)$ Skyrme~\cite{S}
model. Like in that case~\cite{S}, this term serves only the purpose
of fixing the asymptotics, and will be considered only in this limited
context. Moreover, anticipating our conclusions in Subsection~\ref{zero},
namely that only for $\omega=0$ is it possible to construct finite
action solutions, the relevant $V(\phi^5,\omega=0)$ is
\be
\label{pot}
V(\phi^5,\omega=0)=\lambda(1-\phi^5)\ ,
\ee
in analogy with the pion--mass~\cite{S} potential. Notwithstanding, in
Subsection~\ref{pibytwo} we have described a model differing from \re{L}
and characterised by nonzero $\omega$ (in particular
$\omega=\frac{\pi}{2}$), which can support topologically stable finite
action solutions. As will be seen there, the Lagrangian of such a model
does not feature the usual YM term and is therefore not studied further
here.

Our topologically stable finite action solutions are interpreted as
instantons, although the latter are not always characterised by the
second Chern--Pontryagin density, but also by the $S^4\to S^4$ degree
of the $S^4$ valued field (which may or may not be integer).

By adapting the methods formulated in \cite{T} used in establishing the
topological lower bounds given by the degree of the map $S^4\to S^4$,
a suitable such lower bound can be established for the action \re{L},
for all positive values of $\lambda_1 , \lambda_2 , \lambda_3 ,$ and
$\lambda_0$. This will be established in Section~\ref{lowerbounds}. In Section~\ref{sphsymm}, the system \re{L} will be subjected to spherical
symmetry, and thenceforth we will restrict to the study of the ensuing
one dimensional equations numerically. The results of our numerical work
is presented in Section~\ref{numerical}, and Section~\ref{sumdisc} is
devoted to the summary and discussion of our results.

In the work of \cite{nl}, which employs the above model \re{L} with
$\la_1=\la_3=0$, it was found that the action of the main
$SO(4)$ gauged $O(5)$ model decreased with the
Skyrme coupling $\lambda_2$ and exhibited a bifurcation at a value
very close to and above the action of the (pure) Yang--Mills (YM)
instanton~\cite{BPST}. Thus beyond a critical value of the coupling
$\lambda_2$, the system did not support a finite action solution, and
more importantly the action could be made smaller by decreasing
$\lambda_2$. Our main aim in the present work is the verification of
these two properties of the solutions when $\la_1>0$ and $\la_3>0$. We
have found that these properties of the solution persist, namely that
instantons can be constructed for values of $\la_2$ (holding $\la_3$
constant) up to some value $\la_2^{{cr}}$, and for values of $\la_3$
(holding $\la_2$ constant) up to some value $\la_3^{{cr}}$. Also, the
actions of these instantons decrease with $\la_2$ and $\la_3$ respectively
consistently with the Derrick scaling requirement. We find surprisingly
that the action at $\la_2=0$ is nonzero, inspite of the vanishing of
the topological lower bound at that point. We defer discussion of the
possible physical significance of these properties to
Section~\ref{sumdisc}.

Another objective here is to probe the nature of the topological lower
bounds. In \cite{nl}, we were exclusively concerned with lower bounds
stated in terms of the degree of the map of the $S^4$ valued field,
such that the corresponding solutions supported {\em vanishing}
Chern--Pontryagin charge. Here we attempt to construct numerically,
instantons with {\em nonzero} Pontryagin charge when the $S^4$ field
is nontrivial. We do not find such solutions and offer an analytic
argument to support their nonexistence.

The boundary conditions employed for the $S^4$ valued field are 
\be
\label{asym}
\lim_{r\to 0}\phi^5=-1\ ,\qquad \lim_{r\to \infty}\phi^5=\cos \omega\ ,
\ee
but we will be restricting our numerical investigations to the
$\omega=0$ case. For the gauge field we will adopt the vacuum behaviours
\be
\label{vac}
\lim_{r\to 0}A_{\mu}=0\ ,\qquad
\lim_{r\to \infty}A_{\mu}=q g\ \pa_{\mu}g^{-1}\ .
\ee
$q=0$ leads to zero Pontryagin charge. Pure-gauge $q=1$, and
half--pure-gauge $q=\frac{1}{2}$, both lead to nonzero Pontryagin
charges. The half--pure-gauge case $q=\frac{1}{2}$ pertains to
$\omega\neq 0$, which we do not study numerically. The pure-gauge cases
$q=0$ and $q=1$ both pertain to $\omega=0$, and are studied numerically.
It turns out that instantons with nontrivial ($S^4$ valued) matter field
can only be constructed for the $q=0$ case.

\section{Lower bounds}
\label{lowerbounds}

The work in this Section follows very closely that in \cite{T} and that
in Section~\ref{lowerbounds} of \cite{nl}. The analysis in \cite{T,nl}
was adapted to the definition of topological charges presenting lower
bounds on the actions, of systems supporting asymptotics with $\omega=0$
in \re{asym}. Here, we extend this analysis to include values of
$0\le\omega\le\pi$. In the generic case therefore, we would expect
solutions with a {\em fractional} analogue of the Baryon number in 3
dimensions~\cite{BHT}, while in the limiting case~\cite{T,nl}
$\omega=0$ this will be the degree of the map of the $S^5$ valued field
taking an {\em integer} value.

We start with the definition of the winding number density
\be
\varrho_0=\frac{1}{64\pi^2}\vep_{\mu\nu\rho\sigma}\vep^{abcde}
\pa_{\mu}\phi^a\pa_{\nu}\phi^b\pa_{\rho}\phi^c\pa_{\sigma}\phi^d\phi^e\ ,
\label{r0}
\ee
which is inadequate for our purposes here since it is {\em gauge variant},
and its {\em gauge invariant} version
\be
\varrho_G=\frac{1}{64\pi^2}\vep_{\mu\nu\rho\sigma}\vep^{abcde}
D_{\mu}\phi^aD_{\nu}\phi^bD_{\rho}\phi^cD_{\sigma}\phi^d\phi^e\ ,
\label{rG}
\ee
whose volume integral cannot be evaluated by stating the asymptotic
conditions, {\it i.e.} it is not useful as a topological charge density.

The volume integral of \re{r0} can indeed be evaluated by stating the
asymptotic conditions \re{asym}, such that for $\omega=0$ the value of
this integral is integer, and is fractional for $\omega\neq0$. The
normalisation ensures that in the spherically symmetric case this is
the unit charge, or winding number. The task is to define a suitable
density which is {\bf (a)} gauge invariant, and {\bf (b)} its volume
integral equals the volume integral of the density $\varrho_0$, \re{r0}.

To this end we find the relation between the two densities \re{r0} and
\re{rG}, in suitable form, such that {\em all gauge--variant terms
appear as total divergences}. Thus,
\bea
\varrho_G&=&\varrho_0+\pa_{\mu}\left(\phi^5\pa_{\nu}\Omega_{\nu\mu}+
\tilde \Omega_{\mu}\right) \nonumber \\
&&+\frac{3}{64\pi^2}
\vep_{\mu\nu\rho\sigma}\vep^{\alpha\beta\gamma\delta}
\left[\phi^{\alpha}D_{\nu}\phi^{\beta}D_{\mu}\phi^5
F_{\rho\sigma}^{\gamma\delta}-
\frac{1}{8}\phi^5\left(1-\frac{1}{3}(\phi)^2\right)
F_{\mu\nu}^{\alpha\beta}F_{\rho\sigma}^{\gamma\delta}
\right] \label{rel1}
\eea
where
\be
\Omega_{\nu\mu}=\frac{3}{64\pi^2}
\vep_{\mu\nu\rho\sigma}\vep^{\alpha\beta\gamma\delta}
A_{\sigma}\phi^{\alpha}\left(2\pa_{\rho}\phi^{\beta}+
(A_{\rho}\phi)^{\beta}\right)
\label{o1}
\ee
\be
\tilde \Omega_{\mu}=
\frac{3}{128\pi^2}
\vep_{\mu\nu\rho\sigma}\vep^{\alpha\beta\gamma\delta}
\left\{\phi^5\left(1-\frac{1}{3}(\phi)^2\right)A_{\nu}^{\gamma\delta}
\left[\pa_{\rho}A_{\sigma}^{\alpha\beta}-
\frac{2}{3}(A_{\rho}A_{\sigma})^{\alpha\beta}\right]\right\}\ .
\label{o2}
\ee

The volume integral of the (gauge--variant) total divergence terms in
\re{rel1} can, after conversion to two surface integrals, be evaluated
using the asymptotic values \re{asym} and \re{vac}. If these
surface integrals vanished, then the volume integral of the remaining
gauge invariant terms could be expressed in terms of the volume integral
of $\varrho_0$, namely the (integral or fractional) winding number,
leading to the definition of a gauge-invariant topological charge
density.

This can be checked by substituting the spherically symmetric Ansatz
\re{spha} and \re{sphf} in \re{o1} and \re{o2}. One then sees
immediately that the density \re{o1} vanishes asymptotically by symmetry,
while the density \re{o2} yields a non-vanishing contribution to the
corresponding surface integral, which depends on the asymptotic parameter
$\omega$. Thus the definition of the charge density to be given below
is $\omega$-dependent.

Making use of the relation
\be
\frac{1}{4}
\vep_{\mu\nu\rho\sigma}\vep^{\alpha\beta\gamma\delta}
F_{\mu\nu}^{\alpha\beta}F_{\rho\sigma}^{\gamma\delta}=
\vep_{\mu\nu\rho\sigma}\vep^{\alpha\beta\gamma\delta}\pa_{\mu}\left\{
A_{\nu}^{\gamma\delta}\left[\pa_{\rho}A_{\sigma}^{\alpha\beta}-
\frac{2}{3}(A_{\rho}A_{\sigma})^{\alpha\beta}\right]\right\}
\label{euler}
\ee
one can add and subtract the gauge--invariant
density
\[
\cos\omega\left(\cos\omega-\frac{1}{3}\cos^2\omega\right)
\vep_{\mu\nu\rho\sigma}\vep^{\alpha\beta\gamma\delta}
F_{\mu\nu}^{\alpha\beta}F_{\rho\sigma}^{\gamma\delta}
\]
to \re{rel1}, such that the non-vanishing surface contribution of
$\tilde\Omega_{\mu}$ is cancelled. After some rearrangement, the natural
definition for the charge density $\varrho$ pertaining to a system
characterised by the asymptotic parameter $\omega$ is
\bea
\varrho=\varrho_G&-&
\frac{3}{64\pi^2}
\vep_{\mu\nu\rho\sigma}\vep^{\alpha\beta\gamma\delta}
\Bigg\{
\phi^{\alpha}D_{\nu}\phi^{\beta}D_{\mu}\phi^5
F_{\rho\sigma}^{\gamma\delta} \nonumber \\
&-&\frac{1}{8}\left[\phi^5\left(1-\frac{1}{3}(\phi)^2\right)-
\cos\omega\left(\cos\omega-\frac{1}{3}\cos^2\omega\right)
\right]
F_{\mu\nu}^{\alpha\beta}F_{\rho\sigma}^{\gamma\delta}\Bigg\}\ ,
\label{ch1}
\eea
which is the manifestly gauge--invariant definition that is employed in
establishing (Bogomol'nyi like) lower bounds, and which is equivalent to
the definition
\be
\varrho=\varrho_0+\pa_{\mu}\left(\phi^5\pa_{\nu}\Omega_{\nu\mu}+
\hat \Omega_{\mu}\right)\ ,
\label{ch2}
\ee
in which $\Omega_{\nu\mu}$ is defined by \re{o1} while $\hat \Omega_{\mu}$
is
\bea
\hat \Omega_{\mu}=
\frac{3}{128\pi^2}
\vep_{\mu\nu\rho\sigma}\vep^{\alpha\beta\gamma\delta}
\Bigg\{&&\left[\phi^5\left(1-\frac{1}{3}(\phi)^2\right)-
\cos\omega\left(\cos\omega-\frac{1}{3}\cos^2\omega\right)
\right]\times \nonumber \\
&& A_{\nu}^{\gamma\delta}
\left[\pa_{\rho}A_{\sigma}^{\alpha\beta}-
\frac{2}{3}(A_{\rho}A_{\sigma})^{\alpha\beta}\right]\ \ \ \Bigg\}\ .
\label{o3}
\eea
For the spherically symmetric fields \re{spha}-\re{sphf},
$x_{\mu}\hat \Omega_{\mu}$ vanishes
asymptotically, and since we already know that $\Omega_{\nu\mu}$ vanishes
asymptotically, we see that the volume integral of \re{ch2}
equals the (generic fractional) winding number. The (topological
charge) density is gauge--invariant, and its volume integral is just
the winding number of the $S^4$ valued field.

Not surprisingly the definition \re{ch1} for the two most prominent
cases $\varrho(\omega=0)$ and $\varrho(\omega=\frac{\pi}{2})$,
simplifies somewhat. After some manipulations one has
\bea
\varrho(0)&=&\varrho_G+\frac{3}{64\pi^2}
\vep_{\mu\nu\rho\sigma}\vep^{\alpha\beta\gamma\delta}
\left[\phi^5D_{\mu}\phi^{\alpha}D_{\nu}\phi^{\beta}
F_{\rho\sigma}^{\gamma\delta}+\frac{1}{12}\left((\phi^5)^3-1\right)
F_{\mu\nu}^{\alpha\beta}F_{\rho\sigma}^{\gamma\delta}\right]
\label{ch1-zero} \\
\varrho({\pi\over2})&=&\varrho_G+\frac{3}{64\pi^2}
\vep_{\mu\nu\rho\sigma}\vep^{\alpha\beta\gamma\delta}
\phi^5\left[D_{\mu}\phi^{\alpha}D_{\nu}\phi^{\beta}
F_{\rho\sigma}^{\gamma\delta}+\frac{1}{12}(\phi^5)^2
F_{\mu\nu}^{\alpha\beta}F_{\rho\sigma}^{\gamma\delta}\right]
\label{ch1-half}
\eea

In the next two Subsections~\ref{zero} and \ref{pibytwo} we shall analyse
models whose actions are bounded from below by the topological charges
corresponding to \re{ch1-zero}, pertaining to $\omega=0$, and
\re{ch1-half}, pertaining to $\omega=\frac{\pi}{2}$.

\subsection{Model with $\omega=0$}
\label{zero}

The relevant gauge--invariant charge density in this case is
\re{ch1-zero}. The first term, $\varrho_G$ can be reproduced as a
consequence of the inequality
\[
\left|\ka_1 D_{\mu}\phi^a-\frac{\ka_3^3}{3!}\vep_{\mu\nu\rho\sigma}
\vep^{abcde}D_{[\nu}\phi^bD_{[\rho}\phi^cD_{\sigma]]}\phi^d\phi^e
\right|^2\ge 0\ ,
\]
leading to
\be
\label{ineq1}
\ka_1^2|D_{\mu}\phi^a|^2+
\ka_3^6|D_{[\nu}\phi^bD_{[\rho}\phi^cD_{\sigma]]}\phi^d|^2
\ge\ka_1\ka_3^3\varrho_G\ ,
\ee
where $\ka_1$ and $\ka_3$ are constants with the dimension of length.
The next term in \re{ch1-zero},
\[
\vep_{\mu\nu\rho\sigma}\vep^{\al\beta\ga\de}\phi^5D_{\mu}\phi^{\al}
D_{\nu}^{\beta}F_{\rho\si}^{\ga\de}\ ,
\]
is reproduced by the inequality
\[
\left|\ka_0^2F_{\mu\nu}^{\al\beta}-\frac{\ka_2^2}{2!^2}
\vep_{\mu\nu\rho\sigma}\vep^{\al\beta\ga\de}\phi^5D_{[\rho}\phi^{\ga}
D_{\si]}\phi^{\de}\right|^2\ge 0\ ,
\]
expanding which and adding the appropriate positive terms to the left
hand side yields
\be
\label{ineq2}
\ka_0^4|F_{\mu\nu}^{\al\beta}|^2+\ka_2^4|D_{[\mu}\phi^aD_{\nu]}\phi^b|^2
\ge \ka_0^2\ka_2^2\vep_{\mu\nu\rho\sigma}\vep^{\al\beta\ga\de}
\phi^5D_{\rho}\phi^{\ga}D_{\si}\phi^{\de}F_{\mu\nu}^{\al\beta}\ .
\ee
Finally the last term in \re{ch1-zero} can be reproduced by adding the
two inequalities
\bea
(\phi^5)^2\left|\phi^5F_{\mu\nu}^{\al\beta}-
\frac{1}{2!^2}F_{\rho\si}^{\ga\de}\right|^2
&\ge& 0\ , \nonumber \\
\left|F_{\mu\nu}^{\al\beta}+\frac{1}{2!^2}F_{\rho\si}^{\ga\de}\right|^2
&\ge& 0\ , \nonumber
\eea
and then adding suitable positive quantities to the right hand side to
yield
\be
\label{ineq3}
\bar \ka_4^4|F_{\mu\nu}^{\al\beta}|^2\ge\frac{1}{3!}\bar \ka_4^4
\vep_{\mu\nu\rho\sigma}\vep^{\al\beta\ga\de}[(\phi^5)^2-1]
F_{\mu\nu}^{\al\beta}F_{\rho\si}^{\ga\de}\ .
\ee
The constants $\ka_0$, in $\ka_1$, $\ka_3$, and $\bar \ka_0$ in
\re{ineq1}, \re{ineq2} and \re{ineq3}, all have the dimension of length.

Adding \re{ineq1}, \re{ineq2} and \re{ineq3}, we end up with an inequality
whose left hand side, up to some redefinitions, is precisely the system
\re{L}, without the potential term $V$. This Lagrangian is bounded
from below by the right hand side,
which will be a topological bound if the latter coincides with the
topological charge density \re{ch1-zero} (up to a constant multiple).
This turns out to be the case, provided that the constants
$\ka_0$, in $\ka_1$, $\ka_3$, and $\bar \ka_0$ satisfy the following
constraints
\be
\label{constraints}
\ka_0^2\ka_2^2=3\ka_1\ka_3^3\ ,\qquad 2\bar \ka_0^4=3\ka_1\ka_3^3\ ,
\ee
with the constant multiplying $\varrho$ \re{ch1-zero}, equal to
$\ka_1\ka_3^3$. Thus the action (before redefining the constants) is 
bounded from below as
\be
\label{bound}
S\ \ge\ \ka_1\ka_3^3\ N\ ,
\ee
where $N$ is the winding number. The action $S$ is the four-volume
integral of the Lagrange density
\bea
\hat{\cal L}&=&(\ka_0^4+\bar\ka_0^4)|F_{\mu \nu}^{[\alpha \beta]}|^2
+\ka_1^2|D_{\mu}\phi^a|^2 \nonumber \\ 
&&+\ka_2^4|D_{\mu}\phi^a \times D_{\nu}\phi^b|^2+\ka_3^6
|D_{\mu}\phi^a \times D_{\nu}\phi^b \times  D_{\rho}\phi^c |^2
\ ,
\label{L1}  
\eea
subject to the restrictions \re{constraints}, which is up to some
redefinitions coincides with \re{L} without the potential term $V$.

It is seen from \re{bound} that the condition that this lower bound be
nontrivial is that neither one of $\ka_1$ and $\ka_3$ should vanish. It
could be thought that this means $\ka_2$ can be set equal to zero
without violating this bound, but from the first member of the constraint
equations \re{constraints} we see that this is impossible. We conclude
therefore that {\bf the lower bound remains valid only as long as all the
constants $(\ka_0,\ka_1,\ka_2,\ka_3,)$, and hence also the couplings
$(\la_0,\la_1,\la_2,\la_3,)$, remain positive and nonzero}.

As will be shown in Section~\ref{sphsymm}, the constants $\la_0$ and
$\la_1$ can be scaled away leaving only two independent coupling
constants $\la_2$ and $\la_3$, both of which have to be positive and
greater than zero if the lower bound \re{bound} is to be preserved.

Before proceeding to the next Subsection, we note that the Lagrangian
\re{L} is not unique in being bounded from below by the topological
charge density \re{ch1-zero}. Rather, it is the simplest system motivated
by the requirements that it features the YM term and the {\it quadratic}
kinetic term of the scalar field. An inspection of the spherically
symmetric restriction of the YM term, eqn \re{f2} below, implies the
finite action condition on the gauge field function $a(r)$
\[
\lim_{r\to\infty} a(r)=\pm 1\ ,
\]
which in the language of the asymptotic conditions \re{vac} means that
$q=0$ and $q=1$, respectively. In the first case, the Pontryagin charge
vanishes, while in the second case it is equal to $1$.

\subsection{Model with $\omega=\frac{\pi}{2}$}
\label{pibytwo}

The relevant gauge--invariant charge density in this case is
\re{ch1-half}. Unlike in the previous Subsection however, here we do not
proceed straightforwardly to construct the simplest density which is
bounded from below by \re{ch1-half}.

The reason is that when $\omega\neq 0$ (as in the case case at hand with
$\omega=\frac{\pi}{2}$) the gauge group $SO(4)$ breaks down to $SO(3)$ at
infinity. This can easily be seen by rotating the asymptotic field
$\phi^{\al}$ ($\al=1,2,3, 4)$ of length
$|\phi^{\al}(\infty)|=\sin \omega$, to the constant vector field along
the $x_4$ axis, by means of an appropriate $SO(4)$ gauge transformation.
The effect of this transformation on the $so(4)$ gauge connection is, that
it develops a line singularity along the $x_4$ axis, and its non-vanishing
components then take their values in the residual $so(3)$. We do not
give the details of the passage to this 'Dirac gauge' here, because this
has been given in detail previously in Refs.~\cite{OT,AOT}, in the
context of the $SO(4)\times U(1)$ gauged Higgs model\footnote{The
analysis here is identical to that in \cite{OT,AOT}, with the $4$
component field $\phi^{\al}$ here replacing the Higgs field
$\Phi=\ga_5\ga_{\mu}\hat x_{\mu}$ of \cite{OT,AOT}. Indeed this is
the case for all $d$-dimensional ($d\ge 3$)$SO(d)$ Higgs models with
$d$-vector Higgs fields\cite{TZ}, of which the most familiar is the
Wu-Yang monopole in $d=3$.}

The relevant information that follows from the preceding discussion is,
that the asymptotic connection field
$\tilde A_{\mu}^{ab}=(\tilde A_i^{ab},
\tilde A_4^{ab})$, $(a=\al ,5)$ in the Dirac gauge decays exactly as
$\frac{1}{r}$, and its only non-vanishing component is
\be
\label{Dirac}
\tilde A_i^{[\al\beta]}=\frac{1}{r(1-\hat x_4)}
(\de_i^{\al}\hat x^{\beta}-\de_j^{\beta}\hat x^{\al})\ ,
\ee
which takes its values in $so(3)$.

It follows that the corresponding asymptotic curvature field has the
Coulomb decay $\frac{1}{r^2}$, and as a consequence the integral of the
YM action density will diverge logarithmically in four dimensions. This
simple fact can also be seen by inspecting the spherically symmetric YM
action density in \re{f2}. Thus, in constructing the density bounded
from below by the topological charge density \re{ch1-half}, it is not
legitimate to employ the usual YM action density.

The remedy is to use instead the YM density constructed from the
antysymmetrised product of two curvature two-forms, namely
\[
|F_{\mu\nu\rho\si}^{abcd}|^2=
|(F_{\mu [\nu}^{\al [\beta}F_{\rho\si]}^{\ga\de]}
+F_{\mu [\nu}^{[\de\ga}F_{\rho\si]}^{\beta]\al})|^2\ .
\]
This term arises naturally in reproducing the last term in the charge
density \re{ch1-half}. To reproduce the second term in the charge density,
it is not legitimate to make use of inequality \re{ineq2} since the latter
features the usual YM density. Given that for the instanton (vacuum)
interpretation of the solution we need to have the quadratic kinetic
term of the scalar field, this necessitates the appearance of the term
\[
|F_{[\mu\nu}^{\al\beta}D_{\la]}\phi^{\ga}|^2\ .
\]
Finally, to reproduce the first term, $\varrho_G$, in \re{ch1-half}, the
most economical option is to adopt the inequality \re{ineq1}. (This
avoids the introduction of the additional and unnecessary term
$|D_{[\mu}\phi^aD_{\nu]}\phi^b|^2$ in the Lagrangian.)

Following the above arguments, we write down the three topological
inequalities corresponding to \re{ineq1}, \re{ineq2}, \re{ineq3} for
the present case with $\omega=\frac{\pi}{2}$. With \re{ineq1} unchanged,
we just give the second two
\bea
\bar\ka_1^2|D_{[\mu}\phi^a|^2
+\bar\ka_3^6|F_{[\mu\nu}^{\al\beta}D_{\la]}\phi^{\ga}|^2
&\ge&3\bar \ka_1\bar\ka_3^3
\vep_{\mu\nu\rho\sigma}\vep^{\al\beta\ga\de}
\phi^5D_{\rho}\phi^{\ga}D_{\si}\phi^{\de}F_{\mu\nu}^{\al\beta}
\label{inequa2} \\
\bar \ka_4^8|F_{\mu\nu\rho\si}^{abcd}|^2+\tau^2(\phi^5)^6
&\ge&\tau\bar \ka_4^4
\frac{3}{2}\vep_{\mu\nu\rho\sigma}\vep^{\al\beta\ga\de}
(\phi^5)^3F_{\mu\nu}^{\al\beta}F_{\rho\si}^{\ga\de}\ ,
\label{inequa3}
\eea
where the constants $\bar \ka_1$, $\bar \ka_3$ and $\bar \ka_4$ all
have the dimension of length, while $\tau$ is dimensionless.

Adding \re{ineq1}, \re{inequa2} and \re{inequa3} results in an inequality
whose right hand side can be identified (up to a numerical factor) with
the topological charge density \re{ch1-half}, provided that
\be
\label{constrts}
\bar\ka_1\bar\ka_3^3=3\ka_1\ka_3^3\ ,
\qquad 6\tau\bar\ka_4^4=\ka_1\ka_3^3\ .
\ee
The resulting topological inequality bounding the action from
below, analogous to \re{bound}, is
\be
\label{bounded}
\tilde S\ge \ka_1\ka_3^3\ N\ ,
\ee
in which the action $\tilde S$ is the four-volume integral of
\bea
\tilde{\cal L}=\bar\ka_4^8|F_{\mu\nu\rho\si}^{abcd}|^2
&+&\bar\ka_3^6|F_{[\mu\nu}^{ab}D_{\la]}\phi^c|^2
+\ka_3^6|D_{\mu}\phi^a \times D_{\nu}\phi^b \times  D_{\rho}\phi^c |^2
\nonumber \\
&+&(\ka_1^2+\bar\ka_1^2)|D_{[\mu}\phi^a|^2+\tau^2(\phi^5)^6\ ,
\label{L2}
\eea
subject to the constraints \re{constrts}. Note that the potential
\re{pott} with $\omega=0$ and $n=6$ appears quite naturally in \re{L2},
and in this case its presence is mandatory if the lower bound on the
action is to be preserved.

Because \re{L2} does not feature the usual YM term besides the $F^4$ term,
it is not likely to be of any physical interest. Hence we do not analyse
it numerically. We note that in the case of the $SO(4)\times U(1)$
Higgs model~\cite{OT,AOT}, which also features the $F^4$ YM term to
the exclusion of the usual $F^2$, it could be argued that at high
energies that system reduced to a conventional $SO(4)\times U(1)$ Higgs
system
\be
\label{Higgs}
\mbox{Tr}\left(\la_2F_{\mu\nu}^2+\la_1D_{\mu}\Phi^2
+\la_0(\Phi^2+\eta^2)^2
\right)\ ,
\ee
where the constant $\eta$ is the VEV of the Higgs field, and all fields
are antihermitian. In other words, the $SO(4)\times U(1)$ Higgs model was
interpreted as the low energy effective action of \re{Higgs}.
Unfortunately, we do not have such an interpretation for the system
\re{L2}.

\section{Spherical symmetry}
\label{sphsymm}

The spherically symmetric Ansatz employed is
\be
\label{spha}
A_{\mu}^{[\alpha \beta ]} =\frac{a(r)-1}{r}
(\delta_{\mu}^{\alpha} \hat x^{\beta}
-\delta_{\mu}^{\beta} \hat x^{\alpha})
\ee
\be
\label{sphf}
\phi^{\alpha} =\sin f(r) \hat x^{\alpha} ,\quad \phi^5 =\cos f(r)
\ee

As explained in Section~\ref{pibytwo}, we will restrict our numerical
analysis to the case of $\omega=0$, and hence give the spherically
symmetric reduction only of the terms in the system \re{L}, or \re{L1},
pertaining to $\omega=0$.

Substituting \re{spha}-\re{sphf} into the component terms of \re{L} we
have, for each term
\bea
|F_{\mu \nu}^{\alpha\beta}|^2
&=&12\left[\left(\frac{a'}{r}\right)^2
+\left(\frac{a^2 -1}{r^2}\right)^2\right]
\label{f2} \\
|D_{\mu}\phi^a|^2&=&f'^2 +3\left(\frac{a^2 \sin^2 f}{r^2}\right)
\label{k2} \\
|D_{\mu}\phi^a \times D_{\nu}\phi^b|^2&=&
12 \left(\frac{a^2 \sin^2 f}{r^2}\right) [f'^2 +\left(\frac{a^2 \sin^2 f}{r^2}\right)]
\label{k4} \\
|D_{\mu}\phi^a \times D_{\nu}\phi^b \times  D_{\rho}\phi^c |^2&=&
36\left(\frac{a^2 \sin^2 f}{r^2}\right)^2
\left[3f'^2 +\left(\frac{a^2 \sin^2 f}{r^2}\right)\right] \label{k6}
\eea
In the following we will study the classical solutions
of the model (\ref{L}) and characterize them
by the classical action $S$ defined  by means of
\be
           S = \frac{1}{8 \pi^2} \int d^4 x {\cal L}
\ee
The reduced 1-dimensional Lagrangian is $r^3$ times the sum, with the
appropriate numerical coefficients in \re{L}, of all the above 4 terms.
We do not display this one-dimensional Lagrangian, nor the ordinary
differential equations that follow. 

The asymptotic values of the function $f(r)$ corresponding to \re{asym}
translate to
\be
\label{fasym}
\lim_{r\to 0}f(r)=\pi\ ,\qquad \lim_{r\to \infty}f(r)=\omega\ ,
\ee
with $\omega=0$, while the asymptotic values of the function $a(r)$
for the cases $q=0$ and $q=1$ translate respectively, to
\bea
\lim_{r\to 0}a(r)&=&1\ , \qquad \lim_{r\to \infty}a(r)\ =\ 1
\label{pont0} \\
\lim_{r\to 0}a(r)&=&1\ , \qquad \lim_{r\to \infty}a(r)\ =\ -1
\label{pont1}\ .
\eea

Let us first point out that the embedded 
charge-one-BPST-instanton solutions~\cite{BPST}
\be
\label{instanton}
a_{BPST}(r) = \frac{k^2-r^2}{k^2+r^2} \ \ \ , \ \ \ f(r) = n \pi\ \
({\rm everywhere})
\ee
($k$ is a real constant, $n$ is an integer)
exist irrespectively of the values of $\lambda_{1,2,3}$ and 
leads to $S_{BPST}= \frac{4}{3}$, corresponding to the action
of the charge-one-instanton solution of the Yang--Mills
theory~\cite{BPST}. Here we are interested in classical solutions with
non constant $f(r)$.

The number of four coupling constants can be reduced to two by using the
following scaling argument. Transforming $r \rightarrow \sigma r$, we have
\begin{equation}
S(\lambda_1,\lambda_2,\lambda_3,\lambda_0) 
 = S(\lambda_1\sigma^2,\lambda_2,\lambda_3\sigma^{-2},\lambda_0)   
 = \lambda_1\sigma^2 
   S\left(1,{\lambda_2\over\lambda_1\sigma^2},
       {\lambda_3\over\lambda_1\sigma^{4}},
       {\lambda_0\over\lambda_1\sigma^2}\right).
\end{equation}
Choosing $\sigma^2 = {\lambda_0\over\lambda_1}$ this gives
\begin{equation}
S(\lambda_1,\lambda_2,\lambda_3,\lambda_0) 
 = \lambda_0
   S\left(1,{\lambda_2\over\lambda_0},
       {\lambda_3\lambda_1\over\lambda_0^2},1\right).  
\end{equation}
In the following we will make use of the above scaling property and
set $\lambda_1 = \lambda_0 = 1$.

\section{Numerical results}
\label{numerical}

We have studied numerically the solutions of the classical equations
associated with (\ref{L}) for the $\omega=0$ model, restricting to the
one dimensional spherically symmetric fields given by \re{f2}-\re{k6}.
Most of the work is carried out with the potential \re{pot} decoupled
i.e. with $\la=0$.

In \cite{nl} the above equations have been studied in detail in the case
$\lambda_1=\lambda_3=0$. Here we want to study the classical solutions
for non-vanishing $\lambda_1, \la_2, \lambda_3$.
Using the standard Derrick scaling argument, it is easily seen that
regular classical solutions will exist only if the coupling constants
$\lambda_1, \lambda_3$ are {\bf both} nonzero. On the other hand,
the topological lower bound \re{bound} derived in the previous Section
states that in addition to $\la_1$ and $\la_0$ (which we have already
set to $\la_0=\la_1=1$ by scaling), both $\la_2$ and $\la_3$
must be positive and nonzero. On the basis of the last statement, there
is no guarantee that the solution persists when $\la_2$ vanishes,
even though this is consistent with the Derrick scaling requirement.

As a result of our numerical studies, we have learnt that with the
asymptotics \re{pont0}, the solution persists when $\la_2$ vanishes. In
this case there remains only one coupling constant to vary, $\la_3$,
which is a simpler case to study and this is presented in
Subsection~\ref{ainf=1.1}. In Subsection~\ref{ainf=1.2}, again with
the asymptotics \re{pont0}, we study the cases where $\la_2$ is varied
for fixed nonzero value of $\la_3$, and also where $\la_3$ is varied
for fixed nonzero value of $\la_2$. These families of solutions all
have Pontryagin charge equal to zero. In Subsection \re{ainf=-1}, we
present the results of our numerical search for solutions with {\it unit}
Pontryagin charge and nontrivial scalar field, with asymptotics
\re{pont1}. The result is negative, and we have supplied an analytic
construction in support of the nonexistence of such instantons.

\subsection{Solutions with $a(0)=a(\infty)=1$ and $\la_2=0$}
\label{ainf=1.1}

With these boundary conditions, the Chern--Pontryagin charge would be
zero and the topological lower bound would be stated in terms of the
degree of the map only.

Integrating the equations for small values of $\lambda_3$ we were able 
to construct solutions with 
\be
a(0)=1 \ \  , \ \  a(\infty)=1 \ \ , \ \ f(0) = \pi \ \ , \ \ f(\infty)=0 
\ee
The profiles of the functions $a,f$ of this solution are presented in
Fig.1 for $\lambda_3 = 0.425$ by the solid lines.
In the limit $\lambda_3=0$ the classical action tends to zero and the
function $a(r)$ tends to a constant~: $a(r)=1$.
When $\lambda_3$ increases, the function $a(r)$ develops a local minimum
(at $r=r_m$) which becomes progressively deeper as indicated in Fig. 2.
The general dependence of $r_m$ on $\lambda_3$ is also reported
on Fig. 2. At the same time the classical action of the solution increases
with $\la_3$, and this is illustrated in Fig. 3.

\vskip 5mm
\begin{figure}[htbp]
\unitlength1cm \hfil
\begin{center}
 \includegraphics[width=8cm,angle=-90]{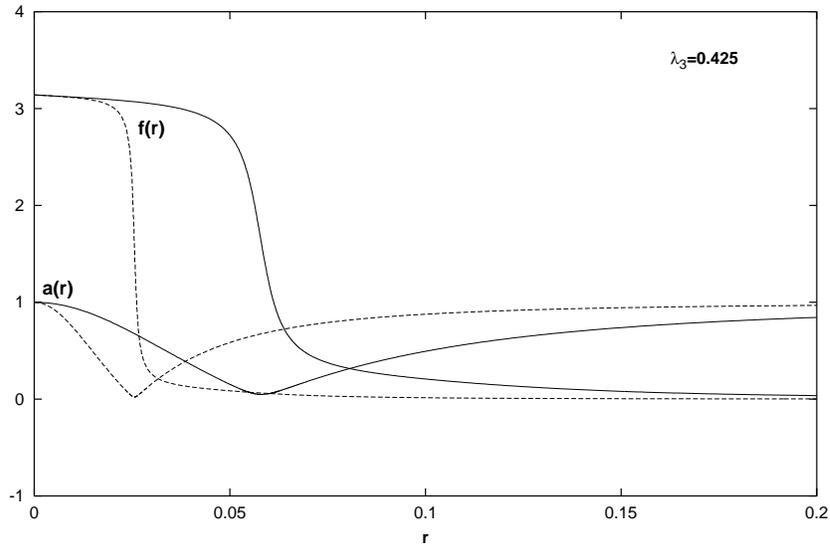}
\end{center}
\caption{The profiles of two solutions $a(r)$ and $f(r)$  as functions of $r$
for $\lambda_2=0$ and $\lambda_3=0.425$ for the first branch (solid line) and 
the second branch (dotted line).}
\end{figure}

\vskip 5mm
\begin{figure}[htbp]
\unitlength1cm \hfil
\begin{center}
 \includegraphics[width=8cm,angle=-90]{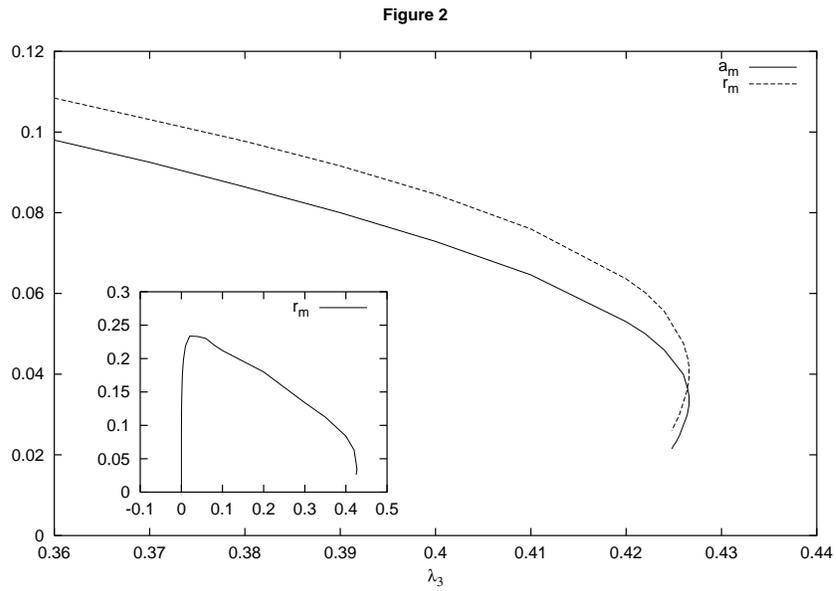}
\end{center}
\caption{$\lambda_3$ dependence of $r_m$ and $a(r_m)$ for the two branches
near the critical value of $\lambda_3$ when $\lambda_2=0.1$.
The global dependence on $r_m$ is displayed in the window.}
\end{figure}

\vskip 5mm
\begin{figure}[htbp]
\unitlength1cm \hfil
\begin{center}
 \includegraphics[width=8cm,angle=-90]{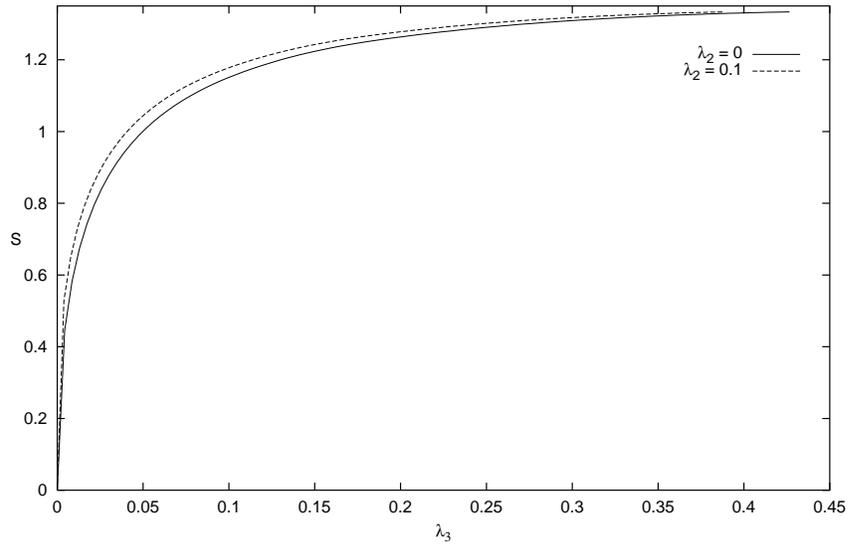}
\end{center}
\caption{$\lambda_3$ dependence of the action for $\lambda_2=0$ (solid line)
and $\lambda_2=0.1$ (dotted line).}
\end{figure}

This situation persists up to a critical value of $\lambda_3$, say 
$\lambda_3^{c}$, and numerically we found 
\be
             \lambda_3^{c} \approx 0.42661 \ .
\ee
Corresponding to this critical value, we find $r_m \approx 0.041$, 
$a(r_m) \approx 0.034$ and $S \approx 1.3345$. In particular the value of
the action is slightly higher than the value $4/3$ corresponding to the
action of the instanton solution of Yang--Mills theory~\cite{BPST}.

In fact, a  large part of this action comes from the Yang-Mills part of
the Lagrangian and the contribution due to the $S^4$ valued matter
 field is rather tiny (less than one percent) because, as indicated
by Fig. 1, the function $f(r)$ becomes
very steep in the region where the function $a(r)$ has its minimum.

We found no solutions for $\lambda_3 > \lambda_3^{c}$; however, 
a careful analysis of the equations strongly suggests that a second
branch of solution exists, as illustrated on Fig. 2. 
As far as the classical action of the two branches is concerned
they terminate into a cusp at $\lambda_3 = \lambda_3^c$, in a 
way very similar to Fig. 9 of \cite{nl}.

As suggested by Fig. 2, 
it is very likely that when $\lambda_3$ decreases to below
$\lambda_3^{c}$ on the second branch, the minimum  $a(r_m)$
has a tendency to approach zero while the derivative $f'(r_m)$
of the function $f(r)$  becomes infinite.
For that reasons, the construction of the classical solution along
this branch is numerically
difficult and we we had to stop it at $\lambda_3 \approx 0.4248$.

Nevertheless, it seems that the profile $a(r)$ of the solution
on the second branch is such that
\be
{\rm lim}_{\lambda_3 \rightarrow \lambda_3^{*}} \ a(r) =
\mid a_{BPST}(r)\mid
\label{limit}
\ee
$a_{BPST}$ being the profile of the 
charge-1 instanton (\ref{instanton})
for an appropriate value of the scaling constant $k$.
The numerical difficulties prevented the evaluation of
$\lambda_3^{*}$ but, according to Fig. 2, one can expect 
$\lambda_3^{*}\approx 0.42$.

The solutions were constructed with the
subroutine COLSYS \cite{colsys} (see Appendix of \cite{bhk}
for a short description) and with a high degree
of accuracy: typically with an error less than $10^{-8}$. 

To finish this Subsection we mention that the pattern of 
solutions presented  above for $\lambda_2 = 0$ seems to persist for
$\lambda_2 > 0$.  For instance, for $\lambda_2  = 1$ we find
$\lambda_3^{c} \approx 0.14$, i.e. a much lower value than 
in the case $\lambda_2 = 0$. More details are given in the next
Subsection.

\subsection{Solutions with $a(1)=a(\infty)=1$ with $\la_2>0, \la_3>0$}
\label{ainf=1.2}

In this Subsection we present numerical results for solutions with the
same asymptotics as in Subsection~\ref{ainf=1.1} above, but {\bf (a)}
holding $\la_2$ fixed at $\la_2=0.1$ and varying $\la_3$, and {\bf (b)}
holding $\la_3$ fixed at $\la_3=0.1$ and varying $\la_2$.

As in Subsection~\ref{ainf=1.1} above, the solutions do not appear to
persist for arbitrarily large $\la_3$ (when $\la_2=0.1$), and arbitrarily
large $\la_2$ (when $\la_3=0.1$). Unlike in Subsection~\ref{ainf=1.1}
however, we have not endeavoured to find accurate critical values for
the $\la_2$ and $\la_3$, respectively. The general features of the
solutions remain unchanged whether or not $\la_2$ vanishes.

\vskip 5mm
\begin{figure}[htbp]
\unitlength1cm \hfil
\begin{center}
 \includegraphics[width=8cm,angle=-90]{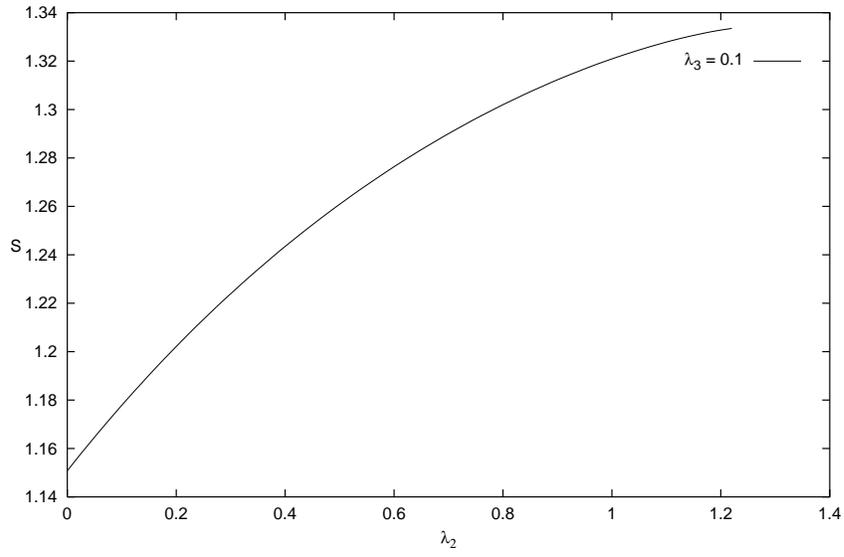}
\end{center}
\caption{$\lambda_2$ dependence of the action for $\lambda_2=0.1$.}
\end{figure}

The action versus $\la_3$ (with $\la_2=0.1$) is plotted on Fig. 3.
The action versus $\la_2$ (with $\la_3=0.1$) is plotted on Fig. 4.
In both cases we note the remarkable feature that the action rises to
just above the BPST instanton action $\frac{4}{3}$. This feature is
shared with the restricted model with $\la_1=\la_3=0$ model studied
in Ref.~\cite{nl}.

\vskip 5mm
\begin{figure}[htbp]
\unitlength1cm \hfil
\begin{center}
 \includegraphics[width=8cm,angle=-90]{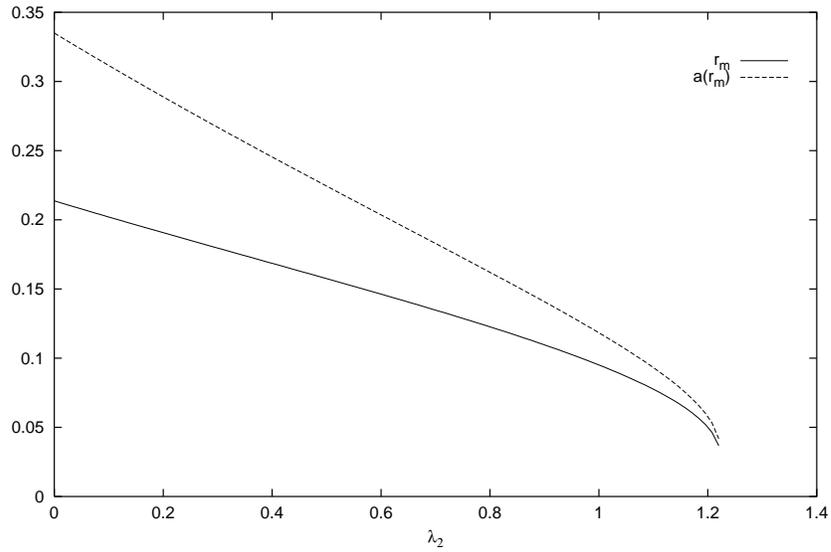}
\end{center}
\caption{$\lambda_2$ dependence of $r_m$ and $a_m$ for $\lambda_3=0.1$.}
\end{figure}

Fig. 5 is the analogue of Fig. 2, where the progress of $r_{min}$ and
$a_{min}$, the position of the minimum and the value of the minimum of
the function $a(r)$, are plotted against increasing $\la_2$ (with
$\la_3=0.1$ fixed). Figs. 2 (resp. Fig. 5) describes the manner in which
the solutions disappear as the value of the $\la_3$ (resp. $\la_2$)
approaches a critical value. (The analysis confirming the existence of
two distinct branches is given only in Fig. 2.)

In addition to the above results, we have made a study of the generic
system \re{L} with the potential \re{pot} included. It turns out that
decoupling the potential \re{pot} results in no appreciable qualitative
changes in the values of the action, for all values of the coupling
constants $\la_2$ and $\la_3$. This property of the present model is
shared by the usual (ungauged) Skyrme model~\cite{S}.

\subsection{Solutions with $a(0)=-a(\infty)=1$}
\label{ainf=-1}

Should solutions of this type exist, their Chern--Pontryagin charge would
be nonzero. As we shall see below, the only such solutions are those
with trivial $S^4$ valued field, i.e. only the pure YM~\cite{BPST}
instantons.

The set of boundary conditions 
\be
a(0)=1 \ \  , \ \  a(\infty)=-1 \ \ , \ \ f(0) = \pi \ \ ,
\ \ f(\infty)=0 \ ,
\label{bc2}
\ee
pertain to {\it unit} Chern--Pontryagin charge for the spherically
symmetric configuration at hand, and provide
a natural alternative to the solution discussed above.

Although we could numerically exhibit regular configurations obeying
these conditions and with an action slightly higher than $4/3$, 
we have not succeeded in constructing a solution of the equations with
these conditions. We used both COLSYS and a relaxation method to find
solutions satisfying the boundary conditions (\ref{bc2}) and all the
numerical results  accumulated  leads us to the formulation of
the following statement: the solutions of the equations of motion
obeying (\ref{bc2}) are constituted by the functions
\be
a(r) = \frac{k^2-r^2}{k^2+r^2} \ \ \ , \ \ \ k \in \R 
\ \ , \ \ f(0) = 0 \ \ , \ \ f(r) = \pi  \ \ , r>0
\ee
considered in the limit $k\rightarrow 0$. The limiting configuration, 
which is determined in terms of step functions, has action $S= 4/3$.

We now give an analytic construction to show that no such smooth solutions
exist. To do this we will
consider a suitably chosen one parameter family of field configurations
and show that in the limit where the parameter vanishes, the field $f$
becomes a step function shrunk near the origin and that the action
reduces to the action of the instanton.

First of all let us write the expression for the action after performing
the scaling $r = \sigma x$ 
\begin{eqnarray} 
S &=& \frac{1}{2}\int \Bigg\{
 \lambda_0 \left(\left(\frac{a_x}{x}\right)^2
+\left(\frac{a^2 -1}{x^2}\right)^2\right)
+\sigma^2 \lambda_1\left(f_x^2
+3\left(\frac{a^2 \sin^2 f}{x^2}\right)\right)
  \nonumber\\ 
&+&\lambda_2 \frac{a^2 \sin^2 f}{x^2} 
    \left( f_x^2 +\frac{a^2 \sin^2 f}{x^2}\right)
 +\frac{\lambda_3}{\sigma^2} \left(\frac{a^2 \sin^2 f}{x^2}\right)^2
  \left(3f_x^2 +\frac{a^2 \sin^2 f}{x^2}\right)\Bigg\} x^3 dx
\end{eqnarray} 
We then consider the following configuration: for $a$ we take the usual 
instanton solution and we call $x_0$ the point where $a(x_0) = 0$. 
For $f$ we take
\begin{eqnarray}
f &=& \pi \hspace{5mm} \mbox{if\qquad} x \le x_0-\frac{\epsilon}{2}\\
f &=& (x_0+\frac{\epsilon}{2}-x) \frac{\pi}{\epsilon} \hspace{5mm}
    \mbox{if\qquad} 
        x_0-\frac{\epsilon}{2} \le x \le x_0+\frac{\epsilon}{2}\\
f &=& 0 \hspace{5mm} \mbox{if\qquad} x \ge x_0+\frac{\epsilon}{2}.
\end{eqnarray}
We then notice that the support of the action density for the first 3 terms 
is the interval $[x_0-\frac{\epsilon}{2},x_0+\frac{\epsilon}{2}]$. 
Moreover in that interval we can write
\begin{equation}
a = K (x-x_0) +O((x-x_0)^2)
\end{equation}
where $K$ is a constant. Defining
\begin{equation}
S_a = \int \frac{\lambda_0}{2} \left((\frac{a_x}{x})^2 
    +(\frac{a^2 -1}{x^2})^2\right) r^3 dr
\end{equation}
we can write
\begin{eqnarray} 
S &=& S_a + \frac{1}{2}\int_{x_0-\epsilon/2}^{x_0+\epsilon/2} \Bigg\{
\sigma^2 \lambda_1 x^3 
   \left(\frac{\pi^2}{\epsilon^2} + \frac{3}{x^2} K^2 (x-x_0)^2\sin^2 f\right) 
\nonumber\\
&+& \lambda_2 x^3 
    \left(\frac{K^2}{x^2} (x-x_0)^2 \sin^2 f 
         (\frac{\pi^2}{\epsilon^2}+\frac{K^2}{x^2} (x-x_0)^2 \sin^2 f ) \right)
\nonumber\\
&+& \frac{\lambda_3}{\sigma^2} x^3 
 \left(\frac{K}{x} (x-x_0)\right)^2
    \left( \frac{3}{\epsilon^2} \pi^2 + \frac{K^2}{x^2}(x-x_0)^2 \sin^2f 
    \right)
\Bigg\}dx
\end{eqnarray} 
We can the replace $\sin f$ by $1$ and perform each of the integrals to the 
lowest order in $\epsilon$ which gives
\begin{eqnarray} 
S&\le& S_a+\sigma^2 \frac{\lambda_1}{2} (\frac{1}{\epsilon}
x_0\pi^2+K^2 x_0 \epsilon^3)+\lambda_1 \sigma^2 O(\epsilon)
+\frac{\lambda_2}{2} K^2 \left(\frac{x_0}{3}\pi^2\epsilon 
+\frac{K^2 \epsilon^5}{5(x_0+\epsilon/2)}\right)\nonumber\\
+ \lambda_2 O(\epsilon^3)
&+&\frac{\lambda_3}{2 \sigma^2} K^4 \left (\frac{3}{(x_0+\epsilon/2)^5}
\epsilon^3+\frac{K^2}{7 (x_0+\epsilon/2)^3} \epsilon^7\right) 
+\frac{\lambda_3}{\sigma^2} O(\epsilon^5)
\end{eqnarray} 
choosing $\sigma = \epsilon $ we have 
\begin{equation}
\lim_{\epsilon \rightarrow 0} S = S_a
\end{equation}
and in that limit the field $f$ becomes singular, showing that
there are no regular solutions with this boundary conditions. This is
indeed what we observed when we tried to compute such solution
numerically.

\section{Summary and discussion}
\label{sumdisc}

The coupling of non-linear sigma models to gauge fields often leads
to sets of classical equations whose solutions obey various 
types of critical phenomena like bifurcations and/or 
pairs of solutions terminating into a cusp.
The classical equations associated with the Lagrangian~\re{L},
in the spherically symmetric Ansatz, are of this type.
These solutions seem to follow the same pattern, irrespectively of the 
different Skyrme terms added, i.e. these patterns seem to
be independent of the dynamical details.

In this paper, we have studied the classical solutions ensuing from the
Lagrangian~\re{L} for three different sets of the two independent
coupling constants $(\la_2,\la_3)$. Inspite of the fact that our
analysis in Section~\ref{lowerbounds} (specifically in
Subsection~\ref{zero}) leads to the establishment of a topological lower
bound on the action provided that {\bf both $\la_2$  and $\la_3$ be
positive}, we have found that in fact solutions persist at $\la_2=0$.
A similar situation arises in the three dimensional Skyrme model
augmented by a sextic Skyrme term. In that case too, when the usual
(quartic) Skyrme term is decoupled, thus invalidating the topological
lower bound, the solution persists~\cite{FP} notwithstanding.
 This most probably means that our (Bogomol'nyi type) analysis in
Section~\ref{lowerbounds} is not far reaching enough for the model at
hand. For example, in the case of Hopf solitons, there exists no
Bogomol'nyi type lower bound on the energy, but instead one finds that
a bound nevertheless can be established in terms of Sobolev type
inequalities~\cite{W}. We have not carried out the appropriate analysis
here, but expect that this can be done. Accordingly we have treated the
simplified model \re{L} with $\la_2=0$ as legitimate and have carried out
the detailed analysis of exhibiting the cusp structure alluded to in the
previous paragraph, for that model in Subsection~\ref{ainf=1.1}, which we
summarise in the next paragraphs.

For the $\la_2=0$ model, clearly the two branches of solutions (the ones
with non trivial $S^4$ valued matter field) terminate into a cusp at
$\lambda_3 = \lambda_3^{c}$. This is a typical situation met in
catastrophy theory.

The spherically symmetric (BPST) instanton of the pure Yang--Mills
theory~\cite{BPST} plays a major role and behaves as an attractor 
(at least when one of the coupling constants approaches a certain value) 
of the solution which excites both the matter and the gauge fields.
It is very likely that the second branch of solutions 
bifurcates from the BPST-branch at the critical value $\lambda_3
=\lambda_3^* < \lambda_3^c$. However, due to the absolute value
in the limit (\ref{limit}), the bifurcating solution does not
occur in a standard, i.e. continuous way.

The qualitative features of the instanton of the $\la_2=0$ model just
described were confirmed also in the generic model with non-vanishing
$\la_2$ and $\la_3$, in Subsections~\ref{ainf=1.2}, where the cusp
resulting from the existence of two distinct branches was not explicitly
constructed.

In Subsection \ref{ainf=-1}, we verified that there existed no instantons
with {\it unit} Pontryagin charge in this model, irrespective of the
value of $S^4\to S^4$ winding number. This is important since it tells
us that the zero Pontryagin charge instantons of this model are not the
analogues of the triangle anomaly, and hence that the nonperturbative
quasiclassical effects they describe must be given a new physical
interpretation. (We return to this in the last paragraph.)

Before alluding to the possible physical relevance of
the model, we note that non-vanishing Pontryagin charge instantons can
readily be constructed by changing the model to feature a symmetry
breaking potential \re{pott} as opposed to \re{pot}. We have
presented the simplest such model in Subsection~\ref{pibytwo}, but did
not carry out a numerical study in that case because the model involved
was rather remote from known physically relevant models.

In short, we have seen that the system consisting of the interacting YM
and $O(5)$ sigma models supports instantons with vanishing Pontryagin
charge, which do not describe quasiclassical effects analogous to the
triangle anomaly, but which have the $S^4\to S^4$ winding number as the
topological charge. Besides, the gauge group for this model is not that
of the Standard Model. On the other hand, it is quite straightforward to
construct an $O(5)$ model interacting with the $SO(3)\times SO(2)$ YM
system that supports such instantons, by adapting the analysis of
Subsection~\ref{lowerbounds} to that case. (That remains a future
project.) Moreover, the number of independent $S^4$ valued fields is
equal to four, just as the number of the Higgs doublet in the Standard
Model, thus an $SO(3)\times SO(2)$ gauged $O(5)$ model added to the
Standard Model, could be regarded as a complicated low energy version of
the latter, whose (axially symmetric) instantons can describe new
nonperturbative effects. In this sense, the $SO(4)$ gauged $O(5)$ model
studied here can be regarded as a prototype of a physically more relevant
model.

\medskip
\medskip

\noindent
{\bf Acknowledgements}: This work was carried out in the framework of
the TMR project TMR/ERBFMRXCT960012, and of Enterprise--Ireland project
IC/2001/073.
\medskip

\medskip

\newpage
\begin{small}

\end{small}
\medskip
\medskip

\end{document}